\documentclass[prl,twocolumn]{revtex4}% Physical Review Letters
\usepackage{graphicx}% Include figure files
\usepackage{dcolumn}% Align table columns on decimal point
\usepackage{bm}% bold math
\usepackage{multirow}% multirow
\usepackage{amsmath,amssymb}% gtrless
\usepackage{threeparttable}% table footnote

\begin{document}

\title{Orbital-selective electronic modes induced by doping and originating from distinct spin excitations: an essential feature of orbital-selective Mott transition}

\author{Masanori Kohno}
\email{KOHNO.Masanori@nims.go.jp}
\affiliation{Research Center for Materials Nanoarchitectonics, National Institute for Materials Science, Tsukuba 305-0003, Japan}

\date{\today}

\begin{abstract}
The nature of the orbital-selective Mott transition (OSMT) remains elusive. This study shows that, by doping an orbitally degenerate spin-1 antiferromagnetic insulator in the Kanamori-Hubbard model, electronic modes emerge in the band gap in an orbital-selective manner, exhibiting momentum-shifted spin-mode dispersion relations. Electronic modes originating from conventional spin excitations, including the low-energy one, appear in the doped orbital, while those from inter-orbital spin excitations appear in the undoped orbital. These emergent modes constitute an essential feature of the OSMT.
\end{abstract}

\maketitle

%\section{Introduction}
{\it Introduction}.--- In conventional band insulators, when the Fermi level enters a band, the system becomes metallic, and only the electrons in that band contribute to conduction. In Mott insulators, this transition is known as the Mott transition. 
However, the true nature of the Mott transition goes far beyond this simple paradigm. Understanding the behavior of electronic states near the Mott transition has been a central challenge in condensed-matter physics. 
\par
The Mott transition has traditionally been considered to be characterized by two mechanisms \cite{ImadaRMP}: (i) vanishing of carrier density via rigid-band filling, as in a conventional band insulator or the antiferromagnetic mean-field theory \cite{SlaterAF}; and (ii) divergence of the effective mass of electronic quasiparticles \cite{BrinkmanRice}, as in the Fermi liquid theory \cite{LandauFL}, i.e., flattening of the dispersion relation at the Fermi level. 
\par
Recent theoretical studies \cite{Kohno1DHub,Kohno2DHub,KohnoDIS,KohnoRPP,KohnoHubLadder,KohnoKLM,KohnoGW,KohnoAF,KohnoSpin,Kohno1DtJ,Kohno2DtJ,KohnoMottT,KohnoTinduced,KohnoDrivenMott}, however, have demonstrated that the Mott transition is better characterized by the loss of spectral weight in the dispersive electronic mode. As the Mott transition is approached, the dispersion reduces to the spin-mode dispersion relation shifted by the Fermi momentum; if the spin excitation is gapless, the electronic mode is also gapless. In contrast to traditional views, the band structure continues to exhibit a dispersive mode without flattening in the metallic phase. This also differs from the interpretations in which emergent electronic modes are gapped despite gapless spin excitations \cite{SakaiImadaPRL,ImadaCofermionPRL,PhillipsRPP,EderOhta2DHub,EderOhtaIPES}. 
%Thus, the Mott transition is accompanied by neither a vanishing carrier density from a rigid band nor a diverging effective mass. 
\par
From the insulating side, the Mott transition can be described as follows \cite{Kohno1DHub,Kohno2DHub,KohnoDIS,KohnoRPP,KohnoHubLadder,KohnoKLM,KohnoGW,KohnoAF,KohnoSpin,Kohno1DtJ,Kohno2DtJ,KohnoMottT,KohnoTinduced,KohnoDrivenMott}: The low-energy spin excitations---whose excitation energies are lower than the charge gap---exist in the Mott insulating phase. Upon doping, the spin-excited states emerge as electronic excitations within the band gap, exhibiting a momentum-shifted spin-mode dispersion relation. The spectral weight of the emergent electronic mode increases in proportion to doping concentration and eventually forms a quasiparticle-like band in the heavily doped regime. 
This behavior reflects spin-charge separation in strongly correlated insulators, i.e., the existence of spin excitations in the energy regime lower than the charge gap, which is absent in conventional band insulators. 
\par
In multi-orbital systems, an insulator--metal transition occurs once the Fermi level enters a band, and the electrons in that orbital become mobile, as in the case of band insulators--- this is referred to as the orbital-selective Mott transition (OSMT) \cite{AnisimovOSMT,GeorgesOSMT}. Analogous to the single-orbital case, it is crucial to clarify the evolution of electronic states near the transition, particularly the relationship between emergent electronic modes upon doping, spin excitations in the insulating phase, and their orbital dependence. Despite extensive studies of the OSMT and electronic states in multi-orbital systems \cite{AnisimovOSMT,GeorgesOSMT,HDDMRG,dopeHDDMFT,undopeHDDMFT,splitHDJDMFT,Hundband}, the nature of the OSMT, a complete understanding of emergent electronic modes, and their underlying connection to spin excitations remain elusive. 
\par
In this Letter, how electronic states in multi-orbital systems evolve upon doping a spin-1 antiferromagnetic insulator is investigated using the non-Abelian dynamical density-matrix renormalization group (DDMRG) method and a theoretical analysis in the one-dimensional (1D) Kanamori-Hubbard model (KHM). In the doped orbital, an electronic mode reflecting the low-energy spin mode--- similarly to the single-orbital case--- emerges, together with an additional electronic mode corresponding to another conventional spin excitation. Even in the undoped orbital, electronic modes reflecting distinct inter-orbital spin excitations emerge. These orbital-selective electronic modes induced by doping constitute an essential feature of the orbital-selective Mott transition.
\par
%\section{Model and methods}
%\subsection{Model, spectral function, and dynamical spin structure factors}
%\subsubsection{Kanamori-Hubbard model}
{\it Model}.--- 
We consider the 1D two-orbital KHM defined by the following Hamiltonian \cite{Kanamori}: 
\begin{align}
{\cal H}=&-\sum_{\langle i,j\rangle,m,\sigma}t_m\left(c_{i,\sigma}^{m\dagger}c_{j,\sigma}^m+{\rm H.c.}\right)-\mu\sum_{i,m}n_i^m\nonumber\\
&+U\sum_{i,m}\left(n_{i,\uparrow}^m-\frac{1}{2}\right)\left(n_{i,\downarrow}^m-\frac{1}{2}\right)\nonumber\\
&+\left(U^{\prime}-\frac{J_{\rm H}}{2}\right)\sum_i\left(n_i^1-1\right)\left(n_i^2-1\right)\nonumber\\
&+J_{\rm H}\sum_i{\bm \eta}_i^1\cdot{\bm \eta}_i^2-J_{\rm H}\sum_i{\bm S}_i^1\cdot{\bm S}_i^2,
\label{eq:Ham}
\end{align}
where $\langle i,j\rangle$ means that sites $i$ and $j$ are nearest neighbors on a chain, and $c_{i,\sigma}^{m\dagger}$ and $n_{i,\sigma}^m$ denote the creation and number operators of an electron with spin $\sigma(=\uparrow,\downarrow)$ on the $m$th orbital ($m=1,2$) at site $i$, respectively. The number operator, spin operator, and $\eta$ operator on the $m$th orbital at site $i$ are denoted by $n_i^m$, ${\bm S}_i^m$, and ${\bm \eta}_i^m$, respectively; 
\begin{equation}
\begin{array}{ll}
n_{i,\sigma}^m=c_{i,\sigma}^{m\dagger}c_{i,\sigma}^m,&n_i^m=\sum_{\sigma}n_{i,\sigma}^m,\\
S_i^{m,z}=\frac{1}{2}\left(n_{i,\uparrow}^m-n_{i,\downarrow}^m\right),&\eta_i^{m,z}=\frac{1}{2}\left(n_i^m-1\right),\\
S^{m,x}_i=\frac{1}{2}\left(S^{m,+}_i+S^{m,-}_i\right),&\eta^{m,x}_i=\frac{1}{2}\left(\eta^{m,+}_i+\eta^{m,-}_i\right),\\
S^{m,y}_i=\frac{1}{2{\rm i}}\left(S^{m,+}_i-S^{m,-}_i\right),&\eta^{m,y}_i=\frac{1}{2{\rm i}}\left(\eta^{m,+}_i-\eta^{m,-}_i\right),\\
S_i^{m,+}=c_{i,\uparrow}^{m\dagger}c_{i,\downarrow}^{m},&\eta_i^{m,+}=(-1)^ic_{i,\uparrow}^{m\dagger}c_{i,\downarrow}^{m\dagger},\\
S_i^{m,-}=c_{i,\downarrow}^{m\dagger}c_{i,\uparrow}^{m},&\eta_i^{m,-}=(-1)^ic_{i,\downarrow}^{m}c_{i,\uparrow}^m,
\end{array}
\label{eq:spinEta}
\end{equation}
where $(-1)^i=+1$ and $-1$ for sites $i$ on the A and B sublattices, respectively. 
%In Eq.~(\ref{eq:Ham}), the atomic orbitals are assumed to be degenerate, and inter-site hopping between different types of atomic orbitals is neglected. We assume orbital rotational invariance, which leads to $U=U^{\prime}+J_{\rm H}$. We also assume $t_1>0$ unless otherwise mentioned. 
We assume $t_1>0$. 
%We also take $t_1(>0)$ as the unit of energy. 
\par
The hole-doping concentration is defined as $\delta=\frac{N_{\rm h}}{N_{\rm s}}$, where $N_{\rm s}$ and $N_{\rm h}$ denote the numbers of sites and doped holes, respectively. At half filling, $\delta=0$. In doped systems, $\mu$ is adjusted so that the ground state has the corresponding $\delta$. At half filling, $\mu$ is set to zero. 
\par
%\subsubsection{Spectral function} % Spectral function
{\it Spectral function and dynamical spin structure factors}.--- 
The spectral function is defined as 
\begin{equation}
A_{\delta}^m(k,\omega)=
\frac{1}{2}\sum_{n,\sigma}\left[
\begin{array}{r}
|\langle n|c_{k,\sigma}^{m\dagger}|{\rm GS}\rangle_{\delta}|^2\delta(\omega-E_n+E_{\rm GS}^{\delta})\\
+|\langle n|c_{k,\sigma}^m|{\rm GS}\rangle_{\delta}|^2\delta(\omega+E_n-E_{\rm GS}^{\delta})
\end{array}\right],
\label{eq:Akw}
\end{equation}
where $|n\rangle$ denotes the $n$th eigenstate of the Hamiltonian with energy $E_n$, and $|{\rm GS}\rangle_{\delta}$ denotes the ground state at doping concentration $\delta$ with energy $E_{\rm GS}^{\delta}$. 
The momentum distribution function is defined as $n_{\delta}^m(k)=\int_{-\infty}^{0}d\omega A_{\delta}^m(k,\omega)$. 
The electronic density of states is defined as $\rho(\omega)=\frac{1}{N_{\rm s}}\sum_k A_{\delta}^m(k,\omega)$, where the summation is take over the Brillouin zone. 
%where $c_{k,\sigma}^{m\dagger}$ is defined as the Fourier transform of $c_{i,\sigma}^{m\dagger}$, 
%\begin{equation}
%\label{eq:cdk}
%c_{k,\sigma}^{m\dagger}=\frac{1}{\sqrt{N_{\rm s}}}\sum_i e^{{\rm i}k r_i}c_{i,\sigma}^{m\dagger}. 
%\end{equation}
\par
%\subsubsection{Dynamical spin structure factors} % Dynamical spin structure factors
The dynamical spin structure factors are defined as 
\begin{align}
\label{eq:Skwpm}
S_{\delta}^{\pm}(k,\omega)&=\sum_{n,\alpha}|\langle n|S^{\pm,\alpha}_k|{\rm GS}\rangle_{\delta}|^2\delta(\omega-E_n+E_{\rm GS}^{\delta}),\\
\label{eq:Tkwpm}
T_{\delta}^{\pm}(k,\omega)&=\sum_{n,\alpha}|\langle n|T^{\pm,\alpha}_k|{\rm GS}\rangle_{\delta}|^2\delta(\omega-E_n+E_{\rm GS}^{\delta})
\end{align}
for $\alpha=x,y,z$, 
and $S^{\pm,\alpha}_k$ and $T^{\pm,\alpha}_k$ are given by the Fourier transforms of $S_i^{\pm,\alpha}$ and $T_i^{\pm,\alpha}$, respectively. The spin operators $S^{\pm,\alpha}_i$ and inter-orbital spin operators $T^{\pm,\alpha}_i$ at site $i$ are defined as 
\begin{equation}
\begin{array}{ll}
S_i^{\pm,\alpha}=\frac{1}{\sqrt{2}}\left(S^{1,\alpha}_i\pm S^{2,\alpha}_i\right)&\text{for}\quad \alpha=x,y,z,\\
S_i^{\pm,+}=S^{\pm,x}_i+{\rm i}S^{\pm,y}_i,&S_i^{\pm,-}=S^{\pm,x}_i-{\rm i}S^{\pm,y}_i,\\
T^{\pm,x}_i=\frac{1}{2}\left(T^{\pm,+}_i+T^{\pm,-}_i\right),&
T^{\pm,+}_i=\frac{1}{\sqrt{2}}\left(c^{1\dagger}_{i,\uparrow}c^2_{i,\downarrow}\pm c^{2\dagger}_{i,\uparrow}c^1_{i,\downarrow}\right),\\
T^{\pm,y}_i=\frac{1}{2{\rm i}}\left(T^{\pm,+}_i-T^{\pm,-}_i\right),&
T^{\pm,-}_i=\frac{1}{\sqrt{2}}\left(c^{1\dagger}_{i,\downarrow}c^2_{i,\uparrow}\pm c^{2\dagger}_{i,\downarrow}c^1_{i,\uparrow}\right),\\
\multicolumn{2}{l}{
T^{\pm,z}_i=\frac{1}{2\sqrt{2}}\left(c^{1\dagger}_{i,\uparrow}c^2_{i,\uparrow}\pm c^{2\dagger}_{i,\uparrow}c^1_{i,\uparrow}-c^{1\dagger}_{i,\downarrow}c^2_{i,\downarrow}\mp c^{2\dagger}_{i,\downarrow}c^1_{i,\downarrow}\right).}
\end{array}
\end{equation}
The Fourier transforms of $S_i^{\pm,\gamma}$ and $T_i^{\pm,\gamma}$ are denoted by $S_k^{\pm,\gamma}$ and $T_k^{\pm,\gamma}$, respectively ($\gamma=+,-,z$). 
%These operators transform as SU(2) operators in spin space, since $(-T^{\pm,+}_i,\sqrt{2}T^{\pm,z}_i,T^{\pm,-}_i)$ behave as rank-1 spherical tensor operators with respect to the spin ${\bm S}_i(={\bm S}^1_i+{\bm S}^2_i)$. 
\par
%\subsection{Methods} % Methods
{\it Method}.--- 
For the numerical calculations, the DDMRG method \cite{DDMRG} in the spin-SU(2) non-Abelian basis \cite{nonAbelianHub} was applied to 60-site clusters with open boundary conditions, retaining 240 density-matrix eigenstates \cite{KohnoDIS,Kohno1DtJ,Kohno2DtJ,KohnoHubLadder,KohnoKLM,KohnoGW,KohnoMottT,KohnoTinduced,KohnoDrivenMott,KohnoRPP}. Calculations were performed for $U/t_1=6$, $U^{\prime}/t_1=3.6$, $J_{\rm H}/t_1=2.4$, and $t_2/t_1=0.5$. Gaussian broadening with a standard deviation of $0.1t_1$ was used. 
\par
\begin{figure} % Fig. Akw
\centering
\includegraphics[width=\linewidth]{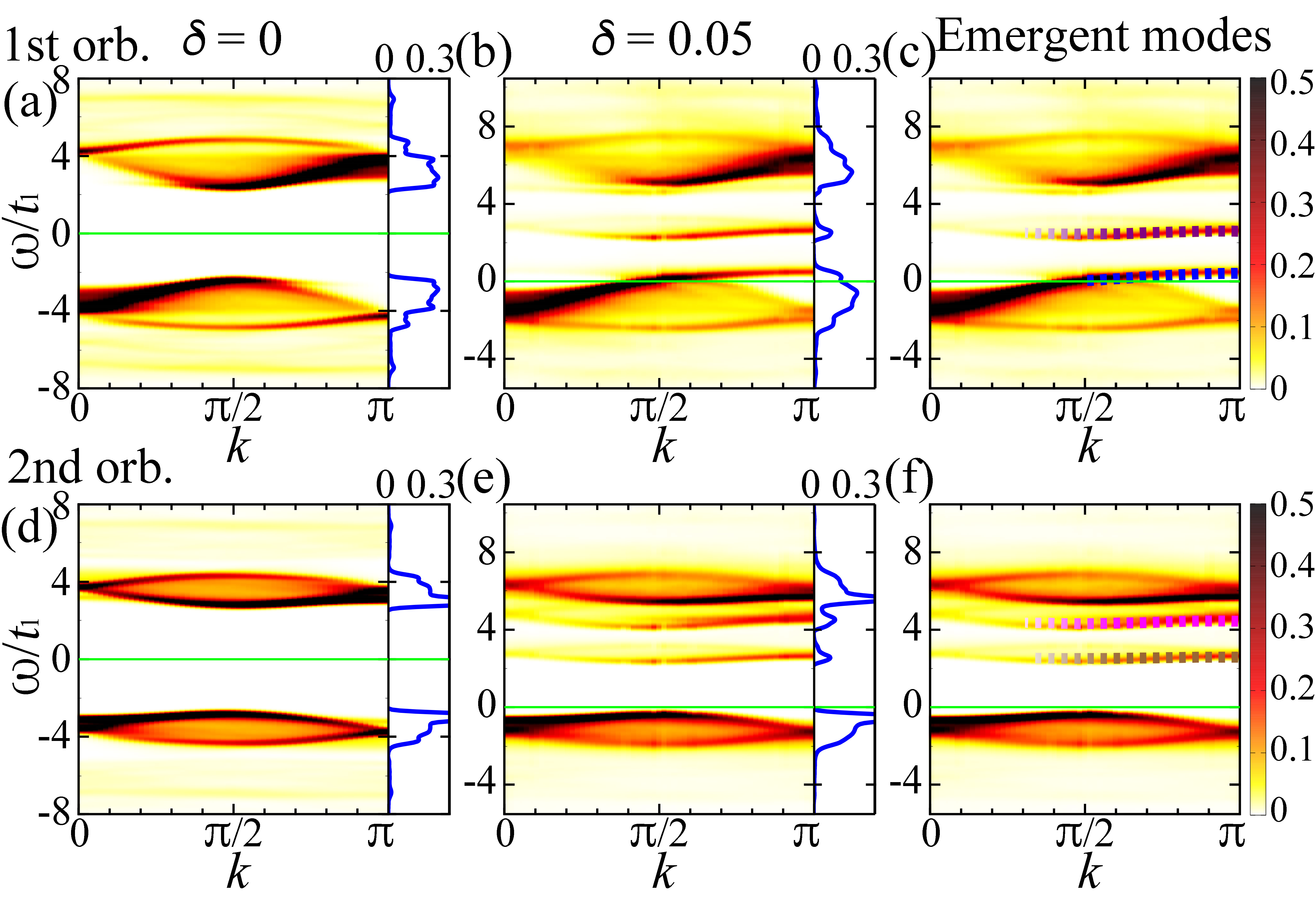}
\caption{$A_{\delta=0}^m(k,\omega)t_1$ [(a), (d)] and $A_{\delta=0.05}^m(k,\omega)t_1$ [(b), (c), (e), (f)] for the 1st orbital ($m=1$) [(a)--(c)] and 2nd orbital ($m=2$) [(d)--(f)]. The blue curves on the right-hand side show $\rho(\omega)t_1$ calculated from $A_{\delta}^m(k,\omega)t_1$ on the left in (a), (b), (d), and (e). The dispersion relations of the doping-induced modes, $\omega=e_{k-\frac{\pi}{2}}^{S^+}$ (dashed blue curve)$, \omega=e_{k-\frac{\pi}{2}}^{S^-}$ (dashed purple curve), $\omega=e_{k-\frac{\pi}{2}}^{T^+}$ (dashed brown curve), and $\omega=e_{k-\frac{\pi}{2}}^{T^-}$ (dashed magenta curve), are plotted in the relevant momentum regimes in (c) and (f). The green lines indicate $\omega=0$.}
\label{fig:Akw}
\end{figure}

%\section{Spectral features} % Spectral features
{\it Spectral features}.--- 
At half filling, the upper Hubbard band (UHB) and the lower Hubbard band (LHB) exist for $\omega>0$ and $\omega<0$, respectively, for each orbital [Figs.~\ref{fig:Akw}(a) and~\ref{fig:Akw}(d)]. The Fermi level is located in the band gap; thus, the system is a Mott insulator. 
\par
Upon hole doping, an electronic mode appears within the band gap from the top of the LHB for the orbital whose upper edge is higher in $\omega$ compared to the other orbital, exhibiting an essentially gapless dispersion relation [Fig.~\ref{fig:Akw}(b)], as found in the Hubbard model \cite{Kohno1DHub}. 
\par
Furthermore, additional electronic modes are induced in the band gap. 
The nature of the emergent electronic modes depends on the orbital; the low-$\omega$ emergent mode appears only in the orbital from which an electron is removed [Fig.~\ref{fig:Akw}(b)], whereas the high-$\omega$ emergent mode just below the UHB primarily appears in the other orbital [Fig.~\ref{fig:Akw}(e)]. The intermediate-$\omega$ emergent modes appear in both orbitals [Figs.~\ref{fig:Akw}(b) and~\ref{fig:Akw}(e)]. 
\par
%\section{Theoretical interpretation}
{\it Interpretations}.--- 
The origins of the emergent modes can be traced back to spin excitations as explained below. 
\begin{figure} % Fig. cartoon
\centering
\includegraphics[width=\linewidth]{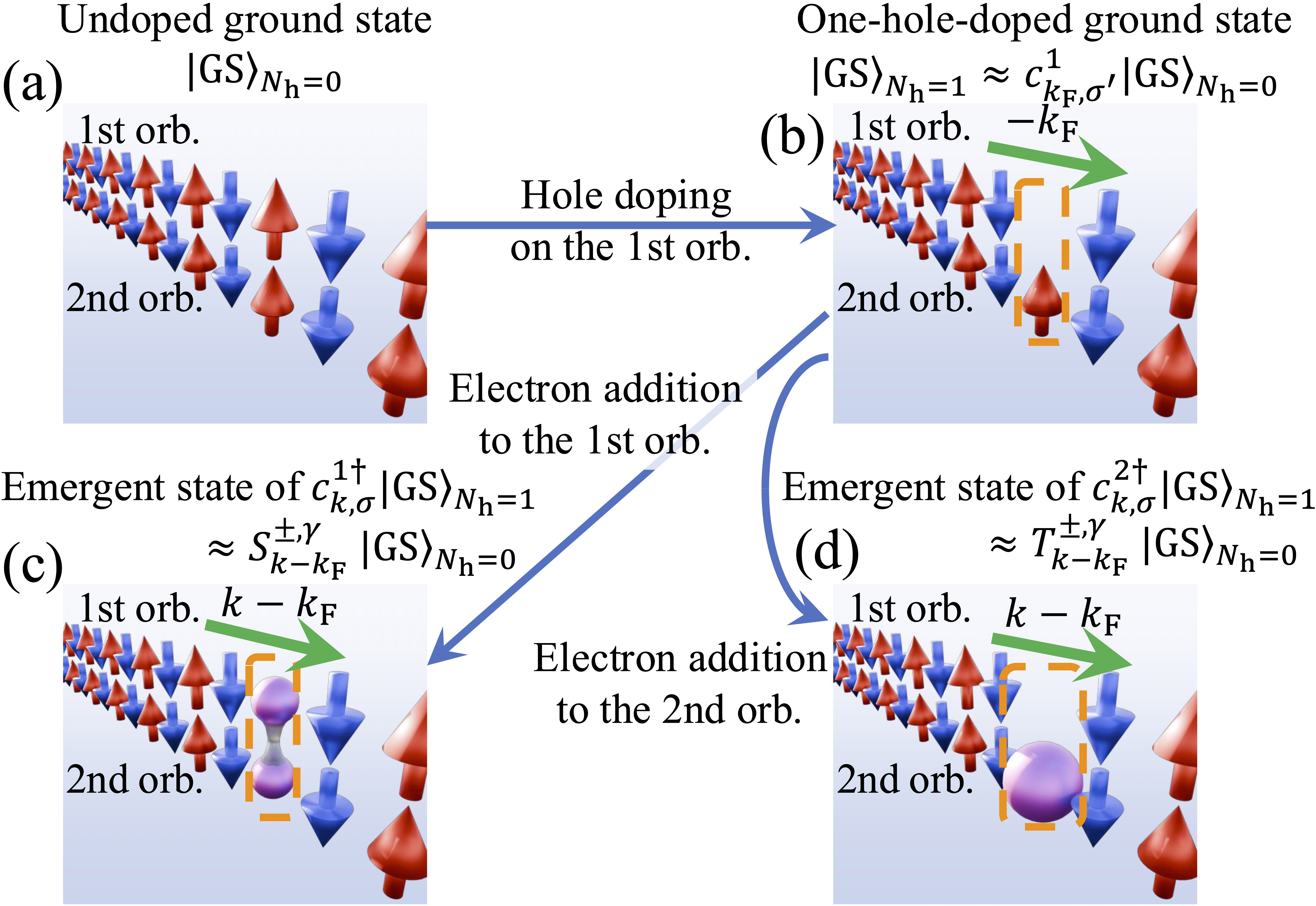}
\caption{Emergence mechanism of spin excited states as electronic excited states upon doping. 
(a) Undoped ground state $|{\rm GS}\rangle_{N_{\rm h}=0}$. (b) One-hole-doped ground state $|{\rm GS}\rangle_{N_{\rm h}=1}$, which has significant overlap with the state obtained by removing an electron with Fermi momentum $k_{\rm F}$ from the first orbital of $|{\rm GS}\rangle_{N_{\rm h}=0}$. The doped site is enclosed by the dashed orange line. 
(c), (d) States obtained by adding an electron with momentum $k$ to the doped orbital [(c)] and undoped orbital [(d)] at the doped site in $|{\rm GS}\rangle_{N_{\rm h}=1}$, which have significant overlap with spin excited states $S_{k-k_{\rm F}}^{\pm,\gamma}|{\rm GS}\rangle_{N_{\rm h}=0}$ [(c)] and $T_{k-k_{\rm F}}^{\pm,\gamma}|{\rm GS}\rangle_{N_{\rm h}=0}$ [(d)], respectively, at half filling. The states enclosed by the dashed orange lines represent a singlet or triplet state with one electron in each orbital [(c)] and a state with no electron in the doped orbital and double occupancy in the undoped orbital [(d)].}
\label{fig:cartoon}
\end{figure}
\par
In the ground state at half filling ($|{\rm GS}\rangle_{N_{\rm h}=0}$), spins are aligned antiferromagnetically with spin-1 at each site [Fig.~\ref{fig:cartoon}(a)]. The momentum and total spin of $|{\rm GS}\rangle_{N_{\rm h}=0}$ are assumed to be zero. 
\par
The one-hole-doped ground state ($|{\rm GS}\rangle_{N_{\rm h}=1}$) has significant overlap with the state obtained by removing an electron from the top of the LHB for the orbital whose upper edge is higher in $\omega$ than that of the other orbital. The momentum at the top of the LHB is denoted by $k_{\rm F}$. Because an electron with momentum $k_{\rm F}$ is removed from $|{\rm GS}\rangle_{N_{\rm h}=0}$, $|{\rm GS}\rangle_{N_{\rm h}=1}$ has momentum $-k_{\rm F}$ and total spin $\frac{1}{2}$ [Fig.~\ref{fig:cartoon}(b)]. 
\par
By adding an electron with momentum $k$ to $|{\rm GS}\rangle_{N_{\rm h}=1}$, electronic excited states with momentum $k-k_{\rm F}$ are obtained. In most cases, the added electron occupies a site that already has one electron in each orbital, which thus contributes to the UHB. However, if an electron is added to the hole-doped site, it contributes to an electronic excitation below the UHB. 
\par
{\it Emergent modes in doped orbital}.--- 
When an electron is added to the doped orbital at the doped site ($i$), the state at that site becomes a spin-triplet state or a spin-singlet state $|{\rm S}\rangle_i$[$=\frac{1}{\sqrt{2}}(|\uparrow,\downarrow\rangle_i-|\downarrow,\uparrow\rangle_i)$] with one electron in each orbital [Fig.~\ref{fig:cartoon}(c)]. These local states are generated by acting $S_i^{+,\gamma}$ and $S_i^{-,\gamma}$ on the spin-triplet state at site $i$ in $|{\rm GS}\rangle_{N_{\rm h}=0}$. 
At other sites, the states are essentially spin-triplet and nearly identical to those in $|{\rm GS}\rangle_{N_{\rm h}=0}$. 
\par
\begin{figure} % Fig. Skw
\centering
\includegraphics[width=\linewidth]{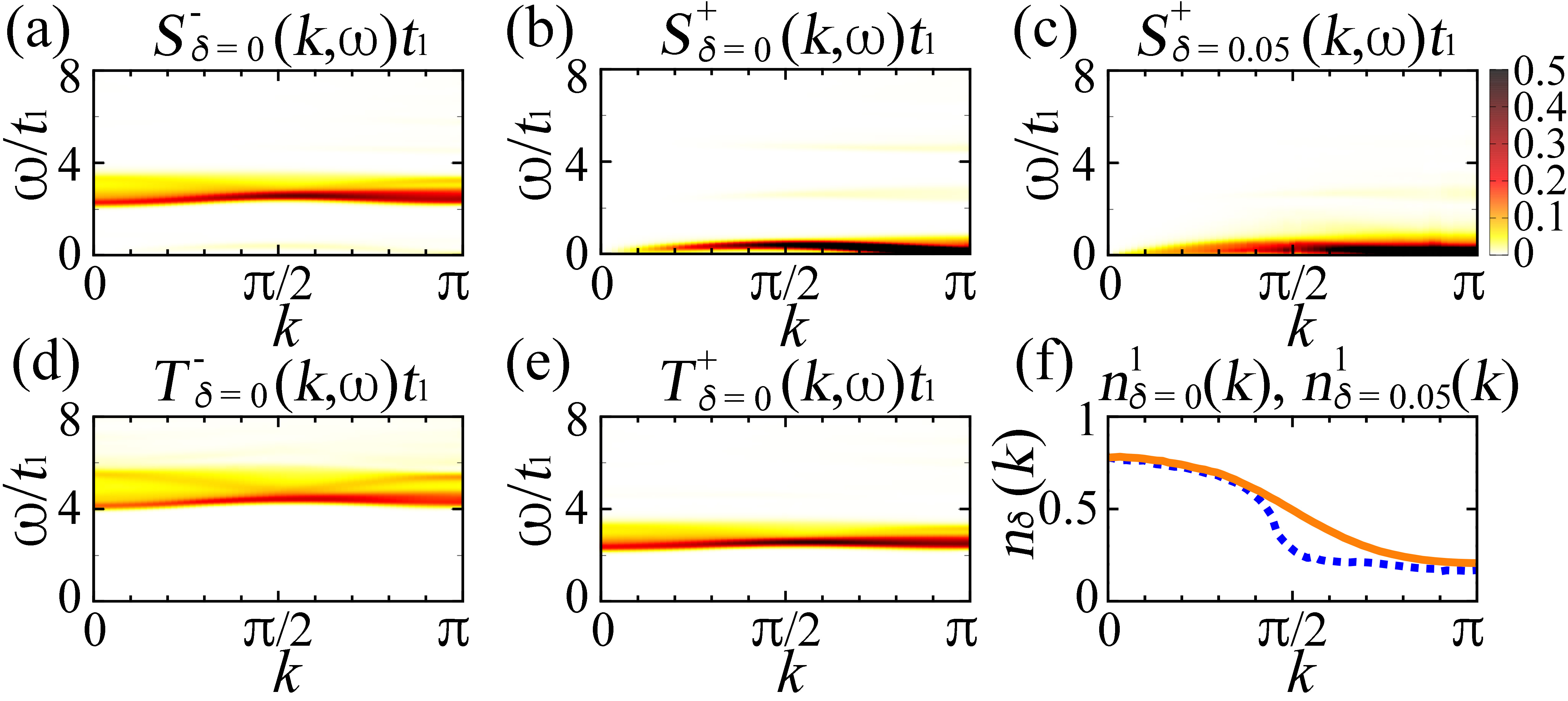}
\caption{(a) $S_{\delta=0}^-(k,\omega)t_1$. (b) $S_{\delta=0}^+(k,\omega)t_1$. (c) $S_{\delta=0.05}^+(k,\omega)t_1$. (d) $T_{\delta=0}^-(k,\omega)t_1$. (e) $T_{\delta=0}^+(k,\omega)t_1$. (f) $n_{\delta=0}^1(k)$ (solid orange curve) and $n_{\delta=0.05}^1(k)$ (dashed blue curve).}
\label{fig:Skw}
\end{figure}

Because the electron-added state has momentum $k-k_{\rm F}$, this state can have significant overlap with the spin excited states $S_{k-k_{\rm F}}^{\pm,\gamma}|{\rm GS}\rangle_{N_{\rm h}=0}$ [Fig.~\ref{fig:cartoon}(c)]. This implies that the spin excited states $S_{k-k_{\rm F}}^{\pm,\gamma}|{\rm GS}\rangle_{N_{\rm h}=0}$ can emerge as electron-addition excited states of the doped orbital in the one-hole-doped system. The dispersion relations of the emergent electronic modes are obtained as 
\begin{equation}
\label{eq:DISS}
\omega=e_{k-k_{\rm F}}^{S^\pm},
\end{equation}
where $e_k^{S^\pm}$ represent the excitation energies of the spin modes in the dynamical spin structure factors $S_{\delta=0}^{\pm}(k,\omega)$ [Eq.~(\ref{eq:Skwpm}); Figs.~\ref{fig:Skw}(a) and~\ref{fig:Skw}(b)]. 
\par
{\it Emergent modes in undoped orbital}.--- 
When an electron is added to the undoped orbital at the doped site ($i$), the state at that site becomes the state with no electron in the doped orbital and double occupancy in the undoped orbital ($|0,\uparrow\downarrow\rangle_i$) [Fig.~\ref{fig:cartoon}(d)]. This local state can be expressed as a combination of the states obtained by acting $T_i^{\pm,\gamma}$ on the spin-triplet state at site $i$ in $|{\rm GS}\rangle_{N_{\rm h}=0}$. 
\par
The electron-added state can have significant overlap with the spin excited states $T_{k-k_{\rm F}}^{\pm,\gamma}|{\rm GS}\rangle_{N_{\rm h}=0}$ [Fig.~\ref{fig:cartoon}(d)], which implies that the spin excited states $T_{k-k_{\rm F}}^{\pm,\gamma}|{\rm GS}\rangle_{N_{\rm h}=0}$ can emerge as electron-addition excited states, exhibiting the dispersion relations expressed as 
\begin{equation}
\label{eq:DIST}
\omega=e_{k-k_{\rm F}}^{T^\pm},
\end{equation}
where $e_k^{T^\pm}$ represent the excitation energies of the spin modes in $T_{\delta=0}^{\pm}(k,\omega)$ [Eq.~(\ref{eq:Tkwpm}); Figs.~\ref{fig:Skw}(d) and~\ref{fig:Skw}(e)]. 
\par
{\it Momentum regimes}.--- 
Because the LHB of the doped orbital in the momentum regime within the Fermi sea remains almost completely filled [Fig.~\ref{fig:Skw}(f)], the electron-addition emergent modes for the doped orbital can appear basically in the momentum regime outside the Fermi sea. 
If the signs of the hopping parameters are the same ($t_2/t_1>0$), the emergent modes for the undoped orbital are also found in the same momentum regime. 
\par
{\it Mode identification}.--- 
The dispersion relations obtained by the above theoretical argument [Eqs.~(\ref{eq:DISS}) and~(\ref{eq:DIST})], together with the spin-mode excitation energies extracted from $S_{\delta=0}^{\pm}(k,\omega)$ and $T_{\delta=0}^{\pm}(k,\omega)$ [Figs.~\ref{fig:Skw}(a),~\ref{fig:Skw}(b),~\ref{fig:Skw}(d), and~\ref{fig:Skw}(e)], can well explain the emergent modes [dashed curves in Figs.~\ref{fig:Akw}(c) and~\ref{fig:Akw}(f)]. 
\par
In the doped orbital, not only does the Fermi level enter the LHB, as in conventional band insulators, but low- and intermediate-energy electronic modes originating from the spin modes in $S_{\delta=0}^{\pm}(k,\omega)$ [Eq.~(\ref{eq:Skwpm}); Figs.~\ref{fig:Skw}(a) and~\ref{fig:Skw}(b)] also emerge from the top of the LHB into the band gap, exhibiting the spin-mode dispersion relations shifted by the Fermi momentum. The emergence of the electronic modes reflecting spin excitations, particularly the low-energy spin mode, is a general and fundamental characteristic of the Mott transition \cite{Kohno1DHub,Kohno2DHub,KohnoDIS,KohnoRPP,KohnoHubLadder,KohnoKLM,KohnoGW,KohnoAF,KohnoSpin,Kohno1DtJ,Kohno2DtJ,KohnoMottT,KohnoTinduced,KohnoDrivenMott}, even in orbitally degenerate systems as shown in this study. 
\par
In the undoped orbital, the emergent modes can be identified with those originating from the spin modes in $T_{\delta=0}^{\pm}(k,\omega)$ [Eq.~(\ref{eq:Tkwpm}); Figs.~\ref{fig:Skw}(d) and~\ref{fig:Skw}(e)]. 
\par
The emergent modes associated with $|{\rm S}\rangle_i$ and $|0,\uparrow\downarrow\rangle_i$ correspond to the states referred to as the Hund band \cite{Hundband} and holon-doublon bound states \cite{HDDMRG,dopeHDDMFT,undopeHDDMFT,splitHDJDMFT}, respectively. This study clarifies their origins, mechanisms, direct connections to the spin excitations, and dispersion relations in a unified manner and in line with manifestations of spin excitations as electronic modes \cite{Kohno1DHub,Kohno2DHub,KohnoDIS,KohnoRPP,KohnoHubLadder,KohnoKLM,KohnoGW,KohnoAF,KohnoSpin,Kohno1DtJ,Kohno2DtJ,KohnoMottT,KohnoTinduced,KohnoDrivenMott}. 
\par
{\it Doping evolution}.--- 
Although the spectral weights of the emergent modes scale as $\mathcal{O}(\frac{1}{N_{\rm s}})$ in the one-hole-doped system, the probability of adding an electron to doped sites increases in proportion to the number of doped sites. This implies that the spectral weights of the emergent electronic modes increase in proportion to doping concentration in the low-doping regime. 
In addition, because the Fermi momentum and the spin excitations in the low-doping regime [Figs.~\ref{fig:Akw}(b) and~\ref{fig:Skw}(c)] are almost the same as those in the low-doping limit and the undoped system, respectively, the dispersion relations of the emergent electronic modes also remain nearly unchanged from the low-doping limit. 
Hence, as doping concentration increases, the emergent electronic modes gradually evolve into well-developed electronic bands; in particular, the low-energy emergent mode in the doped orbital becomes a gapless mode carrying significant spectral weight which may be regarded as a quasiparticle of a Fermi liquid. 
\par
{\it Other emergent modes and generalizations}.--- 
The emergence of electronic modes at nonzero temperatures and in spin- or charge-perturbed nonequilibrium states, the formation of spin-triplet pairs upon doping for small inter-orbital repulsion, and the generalizations to two dimensions, electron doping, and the case with different signs of hopping parameters, as well as the detailed analyses of the doping-induced modes, are described in a separate paper \cite{KohnoKHPRB}. 
\par
{\it Conclusions}.--- 
This study has clarified an essential feature of the OSMT: electronic modes reflecting conventional spin excitations emerge in the band gap from the top of the LHB of the doped orbital upon doping, exhibiting momentum-shifted spin-mode dispersion relations. One of these modes is essentially gapless, reflecting the low-energy spin excitation, and evolves into a quasiparticle-like band with increasing doping. In the undoped orbital, electronic modes reflecting inter-orbital spin excitations emerge in the gap. 
\par
The origins, mechanisms, direct connections to spin excitations, and dispersion relations of these emergent modes are elucidated by theoretical analysis, which consistently explains the emergent modes revealed by numerical calculations for a doped spin-1 antiferromagnetic insulator described by the KHM. 
\par
The OSMT is not simply a phenomenon in which the Fermi level enters a band in an orbital, as in band-insulator--metal transitions, but rather a phenomenon in which an electronic mode reflecting low-energy spin excitations emerges, as in canonical spin-1/2 Mott insulators, only in a doped orbital. 
The emergence of orbital-selective electronic modes, reflecting different types of spin excitations depending on the orbital, is a distinctive characteristic of orbitally degenerate doped large-spin Mott insulators. 
\par
These findings are of broad relevance, as large-spin antiferromagnetic insulators encompass a wide class of transition-metal oxides including Ni, Co, Fe, and Mn-based materials, as well as other correlated electron systems possessing orbital degrees of freedom. Thus, the findings are expected to have broad implications for the understanding of correlated multi-orbital physics and strongly correlated electron systems. 
\par
{\it Acknowledgments}.--- 
This work was supported by JSPS KAKENHI Grant No. JP25K07160 and World Premier International Research Center Initiative (WPI), MEXT, Japan. 
Numerical calculations were partly performed on the Numerical Materials Simulator at NIMS.
%\par
%{\it Data availability}.--- % Data availability
%The data that support the findings of this letter are openly available \cite{dataAvailability}. 

\end{document}